# The Plateau de Bure Neutron Monitor: design, operation and Monte-Carlo simulation

S. Semikh[*], S. Serre, J.L. Autran[#], D. Munteanu, S. Sauze

IM2NP-CNRS, UMR 6242, Aix-Marseille University, Bât. IRPHE, 49 rue Joliot Curie,
BP 146, F-13384 Marseille Cedex 13, France

E. Yakushev, S. Rozov

Dzhelepov Laboratory of Nuclear Problems, JINR, 141980 Dubna, Moscow region, Russia

[*]On leave from Dzhelepov Laboratory of Nuclear Problems

[#] Corresponding author: Phone. +33496139717 – Fax : +33496139709

Email : jean-luc.autran@univ-provence.fr

Abstract - This paper describes the Plateau de Bure Neutron Monitor (PdBNM), an instrument providing continuous ground-level measurements of atmospheric secondary neutron flux resulting from the interaction of primary cosmic rays with the Earth's atmosphere. The detector is installed on the Plateau de Bure (Devoluy mountains, south of France, latitude North 44° 38' 02'', longitude East 5° 54' 26'', altitude 2555 m) as a part of the ASTEP Platform (Altitude Single-event effects Test European Platform), a permanent installation dedicated to the study of the impact of terrestrial natural radiation on microelectronics circuit reliability. The present paper reports the neutron monitor design, its operation since August 2008 and its complete numerical simulation using the Monte Carlo codes GEANT4 and MCNPX. We particularly detail the computation of the neutron monitor detection response function for neutrons, muons, protons and pions, the comparison between GEANT4 and MCNPX numerical results and the evaluation of the PdBNM counting rate a function of both the nature and flux of the incident atmospheric particles.

## 1. Introduction

When a primary cosmic ray strikes atoms in Earth's atmosphere, the collisions may produce one or more new energetic particles called "secondary" cosmic rays. These secondary particles strike other atmospheric atoms producing still more secondary cosmic rays. The whole process is called an atmospheric cascade or extensive air shower. If the primary cosmic ray has enough energy, the nuclear byproducts of the cascade can reach Earth's surface. Such extensive air showers with many particles of different nature (photons, electrons, hadrons, nuclei) arriving on the ground can be detected with a wide variety of particle detectors, including different types of drift chambers, streamer tube detectors, scintillation counters, optical detectors or Geiger tube detectors [1]. Among them, the so-called "neutrons monitors" are especially dedicated to the counting and detection of cosmic rays induced neutrons [1-2]. These neutral atmospheric particles are of great importance in microelectronics, a completely different field than cosmic ray or elementary particle physics, since their interaction with the matter (primarily Silicon) has been identified over many years as a major production mechanism of single-event effects (SEE) in electronics integrated circuits [3-7]. For the most recent deca-nanometers technologies, the impact of other atmospheric particles produced on circuits has been clearly demonstrated (protons) or is still more or less an unexplored question for other exotic particles (pions and charged muons in particular) [3].

In this context and in order to experimentally study the effects of natural atmospheric radiation on microelectronics circuits, we developed and installed in 2005 a permanent test platform in altitude, the Altitude SEE Test European Platform (ASTEP) [8-9]. ASTEP is located in the French Alps on the desert Plateau de Bure (Devoluy Mountains) at 2552m (Latitude North 44° 38' 02'', Longitude East 5° 54' 26''), in a low electromagnetic noise environment, and is hosted by the Institute for Radio-astronomy at Millimeter Wavelengths (IRAM [10]). It has been fully operational since March 2006. From a geomagnetic point-of-view, the ASTEP site is characterized by a cutoff rigidity of 5 GV; the natural neutron flux is approximately 6 times higher that the reference flux measured at New-York City. This value is called "acceleration factor" with respect to the gain that we can expect on the duration of the different experiments for SEE detection performed in altitude instead of at sea-level [3,11].

In 2006 and after suspecting the importance of natural radiation (neutrons) background fluctuations in the interpretation and fine analysis of our experiments [12], we launched the construction of a neutron monitor for the ASTEP platform, precisely to survey on site and in real-time (typically minute per minute) the time variations of the natural atmospheric neutron flux incident on the ASTEP platform. The integration of the instrument was finalized in June 2007; its installation on site was performed in July 2008 after quasi one year of operation and test in Marseille. The instrument, definitively known under the appellation "Plateau de Bure Neutron Monitor" (acronym PdBNM), has been fully operational on ASTEP since July 23, 2008 [13]. In the present paper, we first describe the design and construction of the PdBNM (section 2). In section 3, we report its operation since July 2008 and the observation of detected particle flux fluctuations due to atmospheric pressure variations and solar events. We also detail the experimental determination of the acceleration factor of the ASTEP location with respect to sea-level. In a second part of the paper, we discuss the modeling and numerical simulation of the PdBNM using both GEANT4 and MCNPX Monte Carlo codes. We successively detail in Section 4 the physics models involved in simulations, the comparison methodology between Geant4 and MCNPX results, the input data for realistic particle sources and finally the PdBNM detection responses and its sensitivity to other atmospheric particles (than neutrons). Finally, our conclusions are summarized in section 5.

2. PdBNM design

The Plateau de Bure neutron monitor consists of three $^3$He proportional counter tubes that detect thermal neutrons via the exothermic $^3$He($n$, $p$)$^3$H nuclear reaction (+ 0.764 MeV). The absorption of a neutron by a $^3$He nucleus is generally followed by the emission of charged particles (a triton and a proton) which then are detected by depositing (part of) their energy in the gas and creating a charge cloud in the stopping gas. The greater the energy deposited in the gas by these particles, the larger the number of primary ion pairs, the larger the number of avalanches, and the larger the pulse detected as the output signal by the electronic acquisition chain.

The PdBNM, shown in Fig. 1, is very similar to a standard "3-NM64 neutron monitor" as usually labeled in the literature [1]. Its design follows the recommendations published in [14-15] for the optimization of the apparatus response. The ensemble detector is based on three high pressure (2280 Torr) cylindrical $^3$He detectors; model LND 253109 [16]. These detectors are long tubes (effective length 1828.8 mm) offering a large effective detection volume (3558 cm$^3$) and a very high thermal neutron sensitivity of 1267 counts/nv (LND specifications) [16]. Each tube is surrounded by a 25mm coaxial polyethylene tube which plays the role of a neutron moderator and by 20 coaxial thick (50mm) lead rings serving as secondary neutron producer. All these elements are placed inside a 80mm thick polyethylene box to reject low energy (thermal) neutrons produced in the close vicinity of the instrument. The only geometrical difference between a standard 3-NM64 design and this instrument concerns the exact geometrical shape of the lead rings (rings without lateral extensions, i.e without "ears", but with a flat bottom size) which are directly stacked on the bottom of the polyethylene box [14]. A Canberra electronic detection chain, composed of three charge amplifiers model ACHNP97 and a high voltage source 3200D, was chosen in complement to a Keithley KUSB3116 acquisition module for interfacing the neutron monitor with the control PC. We developed dedicated software under Visual Basic 2008 to control the PdBNM data acquisition as well as to manage and time stamp data using a GPS time acquisition card installed on the same PC. All these operations can be remotely controlled via a VPN connection on internet between ASTEP and the IM2NP laboratory in Marseille. The PdBNM provides in real-time one-minute and uncorrected counting rates for each detection tube, plus temperature, pressure and hygrometry values measured one time per minute at the beginning of the measurement interval. These data are post-processed to provide hourly and monthly averaged values posted on the ASTEP website and available for download [17].

### 3. PdBNM installation and operation

Assembled and firstly operated in Marseille during the year 2007-2008, the PdBNM was transported and definitely installed on the Plateau de Bure in July 2008. Fig. 2 (left) shows the ASTEP building where the instrument is installed. This later has a dedicated local at the first floor of the building with metallic walls (including a 10cm sheet of rockwool), quasi-transparent for high energy neutrons; this building extension was specially constructed to host the neutron monitor in 2007-2008. Fig. 2 (right) schematically shows the position of the neutron monitor inside the building. It is centered with respect to the circular concrete slab (thickness 40cm) forming the floor of the room. We will investigate in Section 4.4.3. by Geant4 simulation the influence of this surrounding building on the neutron monitor detection response. Fig. 3 shows the PdBNM averaged response (one point per hour) from August 1, 2008 to June 21, 2010. This uncorrected response from atmospheric pressure directly gives an image of the neutron flux variation at the ASTEP location; evidencing ~30% variations of this averaged flux at ground level essentially due to atmospheric pressure variations.

During its installation, the PdBNM was used to experimentally determine the acceleration factor (AF) of the ASTEP location with respect to sea-level [11]. With strictly the same setup, two series of data, shown in Fig. 4, were thus recorded in Marseille and on the Plateau de Bure: the difference between the counting rates and barometric coefficients for the two locations allowed us to directly evaluate the acceleration factor of ASTEP with respect to Marseille location, here estimated to 6.7. Taking into account latitude, longitude and altitude corrections for Marseille location with respect to New-York City (the reference place in the world for standardization purposes [11]), the final value of the acceleration factor is AF=6.7x0.94≈6.3. This value is close to 6.2, the average acceleration factor reported in the Annex A of the JEDEC standard JESD89A [11] and close to 5.9, the value given by the Qinetic Atmospheric Radiation Model (QARM) [18-19] for quiet sun activity.

On Fig. 4 is also reported the value of the PdBNM barometric coefficient β used to correct the neutron monitor counting rate from the effect of atmospheric pressure [1]. Data for barometric coefficient calculation have been selected during the period August 2008 – December 2010 for which no disturbance of the interplanetary magnetic field and magnetosphere was reported. The least square method was considered from data of Fig. 4 to obtain the regression coefficient β in the semi-logarithmic representation $Ln(N)$ versus $\exp[-\beta(P - P_0)]$ where N is the hourly neutron monitor counting rate at atmospheric pressure P, β is the so-called barometric coefficient and $P_0$ is the reference atmospheric pressure. Averaged values of β = 0.6642 %/mb has been obtained for the PdBNM with a reference pressure equal to $P_0$ = 749.2 mb at ASTEP location. Under this reference pressure, a counting rate of $N_0$ = 314776 counts.h$^{-1}$ is measured, corresponding to the PdBNM reference counting rate. The magnitude of β also determined is in perfect agreement with typical barometric coefficients values reported in the literature for various instruments [1], suggesting correct response and operation of the instrument. From these values, PdBNM data can be easily corrected from atmospheric pressure using the following well-known transformation:

$$N_{Corrected} = N_{Uncorrected} \exp[-\beta(P - P_0)] \qquad (1)$$

To conclude this section, we would like to illustrate the time dependency of the PdBNM signal and the comparison with the response of other ground neutron monitors for a recent Forbush-effect [20] occurring on February 15-17, 2011. It is the first Forbush decrease in the new solar cycle #24, it means after a very long solar minimum from December 2006. The Sun was active with a X-class flare and many M-class flares during this period [21]. Fig. 5 shows the responses of these different neutron monitors located at Jungfraujoch (Switzerland), Roma (Italy), Athens (Greece), Kerguelen (French islands in south hemisphere) and Plateau de Bure (PdBNM) [22]. The coincidence between the different signals is spectacular during the sudden decrease of the Forbush-effect, as well as the correlation between the signals in terms of amplitude variations before and after this decrease.

### 4. PdBNM Geant4 simulation

In this section, we report in detail the modeling and Monte-Carlo (MC) simulation of the PdBNM detector to study the response of the monitor to the total flux of particles in atmospheric showers, coming from the primary cosmic rays. Our initial objective was to estimate the overall counting rate (and the detection efficiency) of the PdBNM with respect to the total incoming flux and to extract partial contributions from each type of particle (in particular, neutrons) to the overall counting rate. For this purpose, the atmospheric fluxes of all the basic primary particles should be taken into account with corresponding energy spectra and angular dependencies. Following the reference paper by Clem and Dorman in this field [15], we performed simulations for neutrons, muons (μ-, μ+), protons, charged pions (π+, π-), and, in addition, photons, which are considered further as primary particles for PdBNM. Besides, the impact of surrounding to the functioning of PdBNM (concrete floor, building etc.) should also be estimated. As an important intermediate step the evaluation of the PdBNM detection response functions for each type of the incoming particle is required.

Like any other detector system, PdBNM requires an explicit calibration procedure, either with neutron source or with another (calibrated) detector. MC simulation is an important step completing the experimental calibration. In particular, calibration procedure (measurements of the counters spectra) serves also for the correct definition of the detection event in terms of ROI. For PdBNM the explicit calibration is planned for the future, but connected with serious technical difficulties due to its specific location, which increases the role of MC simulation.

As neutron detection event in the neutron counter every single instance of neutron capture on the He-3 nucleus is considered [15]. Generally, the neutron detection event should be defined in terms of response signal level from the proportional counter, belonging to certain ROI, because not all of the events of neutron capture produce the same response due to the wall effects etc. It also allows one to reject possible background and noise. Our experience shows that if ROI is correctly defined then the difference between the adopted definition of detection event and the ROI definition is not huge in terms of the total counting rate, at least it is enough for the estimation of the relative contributions from each type of particle.

Besides, due to the processes of multiple secondary neutron production in the elements of PdBNM (basically, in the lead producer), in this study we have to distinguish between the following quantities: overall counting rate of PdBNM, its total counting rates and detection efficiencies for given particle species. By definition, the efficiency of detecting the given primary particle cannot exceed 1; for the total counting rate (which is always referred to given particle species) it is not so, because for any single primary incoming particle having high enough kinetic energy there is certain probability of multiple neutron production and detection. By overall counting rate we always mean the sum of total counting rates, corresponding to all the components of natural background. The multiplicity of secondary neutrons production in the parts of PdBNM and their detection in counters will be discussed below.

### 4.1. GEANT4 physics models involved in simulations

The list of physical processes employed in the PdBNM detector simulation within GEANT4 toolkit is based on the standard package of physics lists QGSP_BIC_HP [23], significantly modified to include additional interactions of neutrons with polyethylene (PE). Concerning the hadronic interactions, in QGSP group of physics lists the quark gluon string model is applied for high energy interactions of protons, neutrons, pions, kaons and nuclei. The high energy interaction creates an exited nucleus, which is passed to the precompound model describing the nuclear de-excitation. Nuclear capture of negative particles is simulated within the Chiral Invariant Phase Space (CHIPS) model. QGSP_BIC_HP list includes binary cascade for primary protons and neutrons with energies below ~10 GeV, and also uses binary light ion cascade for inelastic interaction of ions up to few GeV/nucleon with matter. In addition this package includes the data driven high precision neutron package (NeutronHP) to transport neutrons below 20 MeV down to thermal energies. But it is well known fact that the moderation of neutrons with kinetic energies below 4 eV in PE should be considered in a special way (see e.g. [24] and references therein). In this low-energy region the scattering of neutrons on the hydrogen nuclei in PE cannot be treated as scattering on free protons due to the possible excitation of vibrational modes in PE molecules. Such collective motion of molecules significantly change the thermal neutron scattering characteristics in PE, so dedicated thermal scattering dataset and model should be included for neutron energies less then 4 eV to allow the correct treatment of neutron moderation and capture processes in the elements of PdBNM. The complete list of the GEANT4 classes for neutrons is given in Table 1 (models and datasets are available since GEANT4.8.2). Another modification of the standard QGSP_BIC_HP list in the current study consists in adding (enabling) the process of muon-nuclear interaction for high energy muons from atmospheric showers. However, it was found that mostly the contribution of this process is not very important here.

### 4.2. Comparison with MCNPX simulation results

To check the adequacy of the adopted physics in GEANT4 it is very informative to compare the results of modeling some test device within different MC codes, e.g. to simulate some simple neutron detector within GEANT4 and MCNPX. Such an independent cross check could reveal possible mistakes in the list of physical interactions. As a test system we consider the employed in PdBNM He-3 neutron counter LND 253109 [16] inserted into PE tube with variable external diameter, as shown on Fig. 6. Let's study the dependence of this detector's response from the energy of mono-energetic neutron flux, taking the neutron source for the simulation as a homogeneous and parallel neutron beam fully illuminating the lateral surface of the system perpendicular to the counter's axis. On Fig. 7 the obtained results are shown for the bare counter and different thicknesses of surrounding PE. Square symbols correspond to MCNPX 2.6.0 evaluation [24], solid lines – to GEANT4.9.1. It is clear that the agreement between different MC codes is good, which confirms the correctness of the adopted physics list in GEANT4.

It is worth mentioning that as detection events in GEANT4 simulation all the instances of neutron capture on He-3 nuclei in active volume of the counter are taken (such events can be easily identified programmatically during tracking). But it is also very useful to calculate within GEANT4 the energy-binned neutron fluence in the active volume of the counter [25] in order to reproduce the procedure of calculating the "Tally F4" in MCNPX (see [26-27] and Appendix for detailed explanation). Convolution of this fluence with neutron capture cross section on He-3 is proportional to the value of response function. Performing this operation in the current study it was explicitly checked that both methods of the test system response evaluation lead to absolutely the same results.

*4.3. Input data for realistic particle sources*

To study the response of PdBNM to natural radiation, one needs realistic energy spectra for each component of the natural background. Energy spectra of particles n, µ-, µ+, p, π+, π-, γ in cosmic-ray induced atmospheric showers, which are employed in corresponding GEANT4 particle sources to simulate real atmospheric fluxes, are available in the literature or on the web as functions of latitude, longitude and altitude. They are obtained from direct measurements and/or from MC simulations. For the neutron flux we use the well-known scalable spectra from [28], which are actually the part of JEDEC Standard JESD89A [11]. Another possibility to obtain all the required spectra is, for example, to refer to QinetiQ atmospheric radiation model (QARM) [18-19] or PARMA model [29-30], which are developed specifically for prediction of the radiation in the atmosphere for a given location and date. In our calculations we mostly employ PARMA model; it is also worth mentioning that in the neutron part its predictions are in a good agreement with JESD89A, as it is seen from the Fig. 8. MC simulations of PdBNM are needed both for Marseille and ASTEP locations. As an example, on Fig. 9 the PARMA differential fluxes for muons and protons are shown at ASTEP.

Another important issue in MC simulation is the strong zenith angular dependence of atmospheric showers. To make GEANT4 primary particle sources more realistic, we introduce in simulations the angular dependence of the primary flux intensity in the form:

$$I(\theta) \sim cos^n(\theta) \qquad (2)$$

where θ is the zenith angle.

Eq. (2) was employed in the GEANT4 built-in General Particle Source [31], allowing us to generate the primary particles with such a given angular distribution. For neutrons we adopt n=3.5 [32], and for muons n=2 [33].

*4.4. Simulation results*

*4.4.1. Detection response functions*

In [15] the detection response functions of neutron monitor NM-64 are evaluated for different particle species as total counting rate versus the energy of mono-energetic particle fluxes, arriving at the upper edge of monitor. For this purpose FLUKA toolkit was employed. To perform the same analysis for PdBNM in GEANT4.9.1, let the incident mono-energetic particles be uniformly distributed upon the upper edge of the PdBNM and arriving in the vertical direction, as it is done in [15]. The obtained dependences of the total counting rate from the energy for different particle species are shown on the Fig. 10. Note, that by definition:

$$Total\ counts = \frac{Number\ of\ captured\ neutrons}{Number\ of\ primary\ neutrons} \qquad (3)$$

and for higher energies this value exceeds 1 due to the processes of secondary neutron production with high multiplicity (see below).

The construction of PdBNM monitor is very similar to the considered in [15] (the difference is basically in geometrical dimensions, but they are still close to each other), so the detection response functions of these monitors can be qualitatively compared. Despite the corresponding curves from [15] cannot be superimposed to Fig. 10, their behavior is in convincing agreement. In particular, all the relative variations are very similar, although for NM-64 the functions are not normalized to the surface area of monitor.

Additionally, the same modeling of PdBNM was performed within MCNPX 2.6.0 [24] for several values of initial energies of neutrons, negative muons and protons (these particle types give dominant contributions to the overall counting rate). MCNPX data points are presented by empty symbols of corresponding colors on the Fig. 10. The models employed are CEM03.01 and LAQGSM03.01 [34-37]. Obtained results demonstrate good agreement, and we can also conclude again that the adopted physical picture of the neutron interactions with matter in GEANT4.9.1 is adequate and close enough to the employed in FLUKA and in MCNPX 2.6.0.

*4.4.2. Secondary neutrons multiplicity*

The monitor PdBNM contains a great amount of lead (about 2 tons), so the contribution of the secondary neutrons production to the total counting rate deserves separate investigation. Within GEANT4 it is not so straightforward to distinguish between the primary and secondary neutron during the tracking, because e.g. elastically scattered primary neutron can formally become the secondary one, but this problem can be solved programmatically. First of all, to verify MC simulation it is quite informative to visualize the distribution of the secondary neutrons production vertices within the volume of the monitor. The example of such a distribution is given at Fig. 11 for the case of primary muons having the energy spectrum obtained within PARMA model [29-30] for the ASTEP conditions and arriving in the vertical direction, which is Z axis (for simplicity we do not reproduce the realistic angular distribution of atmospheric muons in this example). Naturally, the lead tubes produce most of such vertices, although PE walls also contribute.

To analyze the role of secondary neutrons in the detection of primary particles, the GEANT4 simulations were performed both for mono-energetic primary particles and for realistic spectrum. For example, let's consider in details the neutron curve at Fig. 10. For each numerical point of the curve the columns of Table 2 contain the primary neutron kinetic energy $E_{prim}$, detection efficiency, total counting rate, and the ratio of the total number of secondary neutrons produced in the run $N_{sec}$ to the number of primary neutrons $N_{prim} = 10^6$. Any primary particle (neutron, muon, etc.) is considered as *detected* in PdBNM if at least one neutron (either primary or secondary) is captured in He-3 tube in the current event. Neutrons with energies below 1 MeV almost completely reflected by 80 mm PE walls of monitor. Starting from a few MeV, $N_{sec}$ increases very rapidly, finally exceeding $N_{prim}$ for the energies of few tens MeV. The observed difference between efficiency (which is proportional to the number of *detected* primary particles) and total counts is explained by the capture of these additional neutrons in $^3$He tubes within given event.

In principle, to estimate the total counting rate of PdBNM from the atmospheric neutron flux it is sufficient to convolute now the neutron curve from Fig. 10 with the spectrum from Fig. 8. But we prefer to perform direct MC simulation of PdBNM for the realistic neutron source with JEDEC spectrum, thus avoiding, for instance, the errors of numerical interpolation and integration of strongly varying functions over the huge interval of energy.

On Fig. 12 and 13 the obtained for the JEDEC neutron spectrum histograms are presented, which connect the number of secondary neutrons per event and the number of neutron captures in $^3$He counters. Evidently, the neutron production multiplicity can achieve very high values (up to 120 secondary neutrons per event), and the same for the neutron capture multiplicity (up to 14 captured neutrons per event). From this data the contribution of the secondary neutrons to the total counting rate can be analyzed. In fact, the black histogram on Fig. 13 corresponds to the events with single captured neutron, primary or secondary. Its first bin contains the events with no secondary neutrons produced, thus describing the capture of only (moderated) primary neutrons, which corresponds only to 9.7 % of all the captured neutrons in the simulation run. Totally this histogram accumulates 51.1 % of the captured neutrons. The other 48.9 % of neutrons are involved in multiple capture (double, triple, etc.) which distribution is summarized in Table 3 in the following way: the integral from the black histogram on Fig. 13 divided by the total number of captured neutrons in the run is equal to 0.511; the integral from the blue line (events with double neutron captures) multiplied by 2 and divided by the total number of captured neutrons is 0.26 etc.

Totally, within the same simulation run for the secondary neutrons production rate we obtain $N_{sec}/N_{prim} = 1.287$. From Table 2, it is expected that for such a value of secondary neutron production the difference between the detection efficiency and total counting rate can become significant, thus illustrating the importance of secondary neutrons. Let's stress again, that 48.9 % of the neutron counting rate is made by the multiple neutron captures.

The same analysis can be performed for primary particle of any type. It is clear that in detection of the particles other then neutron the role of secondary neutrons production is principal (they are detected as far as they are able to produce secondary neutrons). Considering primary protons and negatively charged muons, for the corresponding curves on Fig. 10 we obtain data shown in Table 4. It is clear, that for protons the secondary neutron production multiplicity is even higher, then for primary neutrons, and the corresponding multiple neutron capture processes in $^3$He are more expressed, as it follows from Table 3. For negative muons, there is a primary energy interval with high secondary neutrons production rate, where the processes of muon moderation and capture in the producer are effective. But for any energy the difference between the detection efficiency and total counting rate does not become as large as for primary neutrons and protons.

### 4.4.3. Impact of the surrounding to PdBNM counting rate

Another important issue for simulation is the impact of surrounding to PdBNM overall counting rate. As described in Section 2, the monitor is placed inside the first floor of the ASTEP building, which possibly distorts the original radiation background. Although the building is relatively light, it contains significant amounts of steel and concrete, which could lead to the secondary neutron production (in steel walls) and neutron reflection (on concrete floor), so the affect of it to the counting rate should be estimated.

The building was included into the total geometry of modeling as shown on Fig. 14 (a). In this simulation, we found that the number of secondary neutrons produced in the system "building + monitor" for the primary protons and neutrons is ~3 times higher than for standalone monitor, and it is ~5 times higher for the primary negative muons. But the resulting impact of this additional neutron flux to the total counting rate is not so significant: it increases total counting rate only in 4.1 % for primary neutrons, 4.5 % for protons and 1.7 % for negative muons. Thus, these extra neutrons are mostly not detected by PdBNM.

To conclude this part, Figs. 14 (b,c) show for illustration different views of a simulated events for five incoming atmospheric neutrons (with energy 100 MeV) interacting with the matter of the PdBNM.

### 4.4.4. Contributions of different particle species to PdBNM counting rate

It is clear that until the experimental calibration of PdBNM is done, only the relative contributions from the different particle species of the natural radiation environment into the overall counting rate of PdBNM monitor can be reliably estimated from MC simulation. Such contributions are directly proportional to the following factors: i) the partial detection efficiency (or total counting rate, see Fig.

10) for the given particle type and ii) its partial flux intensity in the natural conditions depending on the latitude, longitude, altitude, atmospheric pressure, solar cycle etc. To obtain the realistic partial fluxes for the given conditions, we employ the PARMA model and construct a GEANT4 particle source as it is described in section 4.3. Calculated relative contributions to the PdBNM overall counting rate from different types of particles are listed in Table 5 for two completely different locations – Marseille and Plateau de Bure. Higher and lower limits for the percentages arise from the difference between the detection efficiency and total counting rate. It is also worth noticing, that to evaluate total counting rate for primary neutrons it is necessary to consider not only JEDEC part of the neutron flux (from 1 MeV and higher) but also lower part from 1 eV to 1 MeV (available both in PARMA and QARM models), which gives significant contribution due to the very intensive flux in this energy region.

Obviously, PdBNM is quite sensitive to atmospheric protons and negatively charged muons. The latter, being moderated in its materials and captured in nuclei produce a lot of secondary neutrons (it is also shown in Table 4). Sensitivity of PdBNM to other particle species (like electrons and pions) is estimated as very low, either due to the small corresponding value of partial detection efficiency (as for electrons) or low atmospheric partial flux intensity (charged pions).

## 5. Conclusion

In conclusion, this paper summarizes a five years effort to completely develop, install and characterize by simulation a new neutron monitor permanently installed on the Plateau de Bure in French south Alps. The primary interest of the instrument was motivated by the real-time measurement of the atmospheric neutron flux impacting microelectronics experiments deployed in altitude to precisely investigate the impact of natural radiation on electronics. Nevertheless, data obtained with this instrument can be also considered for cosmic rays investigations equally to other monitors installed around the world. Almost three years of continuous operation demonstrated high stability and reliability of the instrument which will soon be integrated into the neutron monitor database (NMDB) network for real-time data accessibility on the web.

Our modeling and numerical simulation work, performed with two different Monte-Carlo codes, GEANT4 and MCNPX, allowed us to obtain the neutron monitor detection response functions for neutrons, muons, protons and pions and their respective contribution into the overall counting rate of the instrument considering realistic atmospheric particle sources. We highlighted a relative importance of proton and negative muon contribution in the monitor response (for muons - especially at sea-level), and a negligible impact of the ASTEP surrounding building on the counting rate. Finally, we carefully characterized the secondary neutron multiplicity processes, essential to understand the physics of this neutron monitor. A future work should be to modify the electronic acquisition chain of the instrument to experimentally characterize such a neutron multiplicity on the electrical signals delivered by the detection tubes and to perform an experimental calibration of the PdBNM.

**Acknowledgments**

The idea to install a Neutron Monitor on the ASTEP platform came after the RADECS Conference organized in Glyfada (Athens) in September 2006 and after the visit of the Athens Neutron Monitor Data Processing Center (ANMODAP) organized in the framework of the IEEE-RADECS SECSTAN Working Group (Susceptibility of Electronic Component and Systems To Atmospheric Neutrons). The authors would like to particularly thank our fellow colleagues Dr. Jean-Luc Leray (CEA) who initiated this legendary visit and Dr. Helen Mavromichalaki (Athens University), principal investigator and scientific leader of the ANMODAP, for her support and continuous encouragement. The logistical support of the Institute for Radio-astronomy at Millimeter Wavelengths (IRAM) is also gratefully acknowledged. Special thanks are due to Bertrand Gautier and Cécile Di Léone for their help. The first author, S. Semikh, would like to specially acknowledge the kind hospitality of the IM2NP laboratory and the group headed by J.L. Autran in conducting this research. The neutron monitor construction was supported in part by the MEDEA+ Project #2A704 ROBIN and by the French Ministry of Economy, Finances and Industry under research conventions #052930215 and #062930210.

**Appendix. MCNPX simulation of neutron flux and reaction rate in the $^3$He counter with Tally F4: fluence response calculation**

In order to determine the fluence response by MCNPX calculation, it is assumed that the number of $^3He(n,p)t$ reactions in the sensitive cell ($V_{s.c.}$) of the proportional counter is correlated to the reading of the counter (detector signal).

The fluence response is determined using an estimation, or *tally*, of the neutron average flux (integrated over time) on the cell containing the sensitive gaseous volume of detection. This tally is defined by the "F4" physic card of the MCNPX code and consists in obtaining the so-called *track length estimate of fluence* $\Phi_j$ through the sensitive cell. It is proportional to the sum of those $K$ path lengths $l_k$ of neutrons having the energy $E_j$ that pass through the sensitive counter volume,

$$\Phi_j \propto \frac{1}{V_{s.c.}} \sum_{k=1}^{K} l_k(E_j) \tag{A1}$$

Finally, $\Phi_j$ is determined by the tally F4 as the area density of the number of neutrons (in cm$^{-2}$) normalized to one source particle passing through the sensitive cell.

The fluence response $R_d(E_n)$ in counts per neutron per cm$^2$ for the cylinder of diameter d, uniformly irradiated by a "plane parallel" and mono-energetic neutron beam with incident energy $E_n$ is then given by

$$R_d(E_n) = \sum_j \Phi_j \, a_s \, n_{He} \, V_{s.c.} \, \sigma(E_j) \tag{A2}$$

with $a_s$ - area of neutron source (in cm$^2$), $n_{He}$ - atomic density of the $^3He$ (in cm$^{-3}$), $V_{s.c.}$ - volume of the counter sensitive cell (in cm$^{-3}$), $\sigma(E_j)$ - cross section of reaction $^3He(n,p)t$ for neutron energy $E_j$ (in barn).

FIGURE AND TABLE CAPTIONS

**Figure 1.** 3D schematics (top) and detailed view (bottom) of the Plateau de Bure Neutron Monitor (PdBNM). Dimensions on the draw are in mm. The instrument is placed on a concrete thick (40 cm) floor.

**Figure 2.** <u>Left</u>: External view of the ASTEP building showing the metal-walled room on the first floor which is installed the Plateau de Bure Neutron Monitor. <u>Right</u>: ROOT screenshot showing the Geant4 modeling of the neutron monitor and its implantation in the ASTEP first floor room.

**Figure 3.** Plateau de Bure Neutron Monitor response recorded from August 1, 2008 to March 31, 2011. Data are uncorrected from atmospheric pressure and averaged over one hour. ~30% variations in neutron flux are evidenced, with several peaks ($>3.8\times10^5$ counts/h) corresponding to the passage of severe atmospheric depressions (the highest peak correspond to the Klaus storm on January 25, 2009).

**Figure 4.** Experimental determination of the ASTEP acceleration factor (AF) from the barometric response of the neutron monitor successively installed in Marseille (2007-2008) and on the Plateau de Bure since July 2008. Experimental clouds correspond to one month recording (one point per hour).

**Figure 5.** Comparison between pressure-corrected signals from five neutrons monitors during the Forbush effect observed on February 15-17, 2011. Data from Jungfraujoch, Roma, Athens and Kerguelen neutron monitors are online available [22]. Courtoisy from the Monitor DataBase (NMDB).

**Figure 6.** Front and side view of the test device for comparative modeling between GEANT4 and MCNPX codes. The He-3 neutron counter model LND 253109 with a 2 inches diameter is inserted into a polyethylene tube with variable external diameter.

**Figure 7.** Response functions of the system defined in Fig. 6 for a bare counter and different external polyethylene tube diameters.

**Figure 8.** Comparison between JEDEC and PARMA atmospheric neutron spectra given for the reference location corresponding to New York City (NYC). After Refs. [11] and [30].

**Figure 9.** Differential fluxes of muons and protons given by the PARMA model for ASTEP conditions.

**Figure 10.** Plateau de Bure neutron monitor detection responses for neutrons, protons, muons and pions. Filled symbols and lines corresponds to GEANT4, large empty symbols corresponds to MCNPX (CEM and LAQGSM models). For any given kinetic energy, the simulation run consists of $10^6$ primary particles (events).

**Figure 11.** 2-dimensional histograms corresponding to the projections of PdBNM to coordinate planes (front and side views) and describing distribution of the secondary neutrons production vertices from incoming muon flux.

**Figure 12.** Two-dimensional histogram connecting the number of secondary neutrons per event and the number of neutron captures in He-3 counters for the JEDEC primary neutron spectrum.

**Figure 13.** Projection of the 2d-histogram at Fig. 12. Each curve corresponds to the fixed number of neutron captures in event.

**Figure 14.** ROOT screenshots of GEANT4 simulations showing the tracks of primary and secondary particles (green lines – neutrons, yellow – gammas, red – electrons) for different views of the instrument: a) top view including the ASTEP surrounding building (first floor); b) face view of the instrument placed on the concrete floor of the building first floor; c) detailed view at the level of the polyethylene box of the neutron monitor.

**Table 1.** List of the considered GEANT4 classes in the simulation flow for the description of neutron interactions.

**Table 2.** Dependence of the secondary neutrons production rate on the primary neutrons energy.

**Table 3.** Contributions to the total counting rate for the given particle type from the events with different multiplicities of neutron captures in $^3$He counters.

**Table 4.** Dependence of the secondary neutrons production rate on the primary protons and negative muons energy.

**Table 5.** Relative contributions to the PdBNM overall counting rate of the different particle species. For the simulation of the instrument response at ASTEP location, the impact of the building is taken into account.

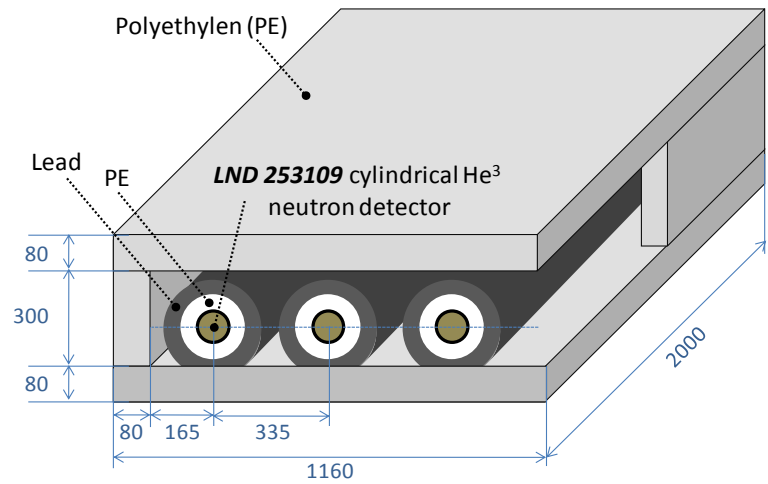
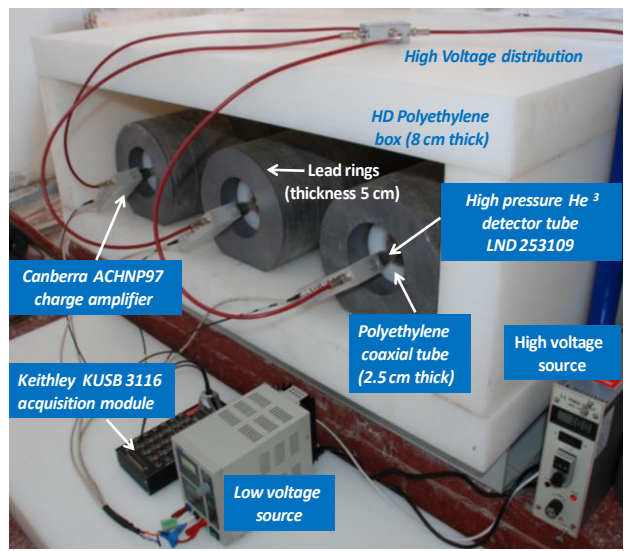

**Figure 1**

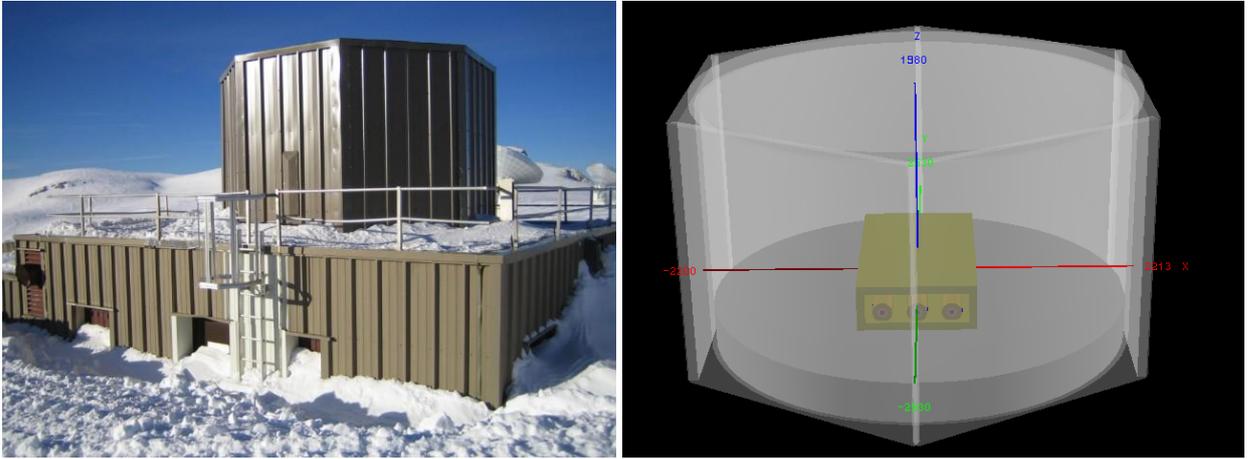

**Figure 2**

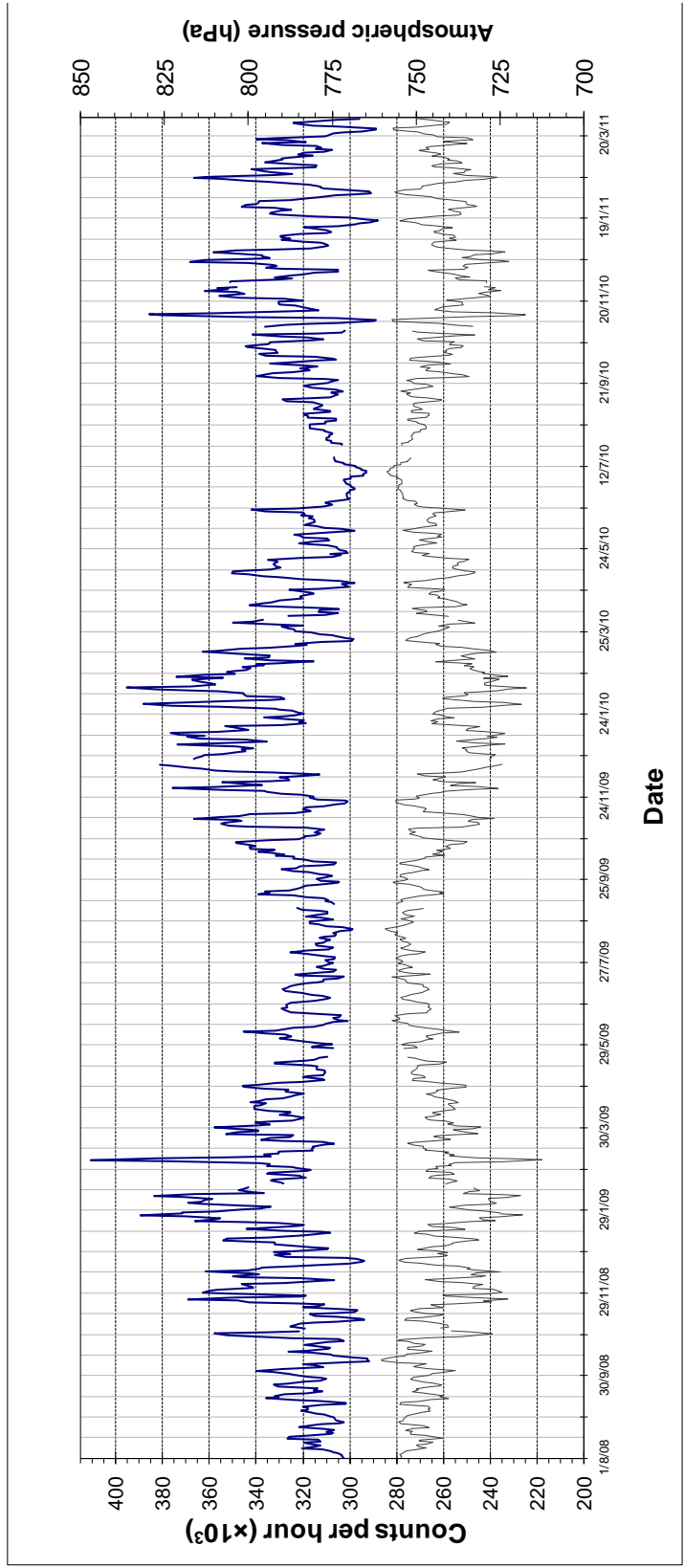

**Figure 3**

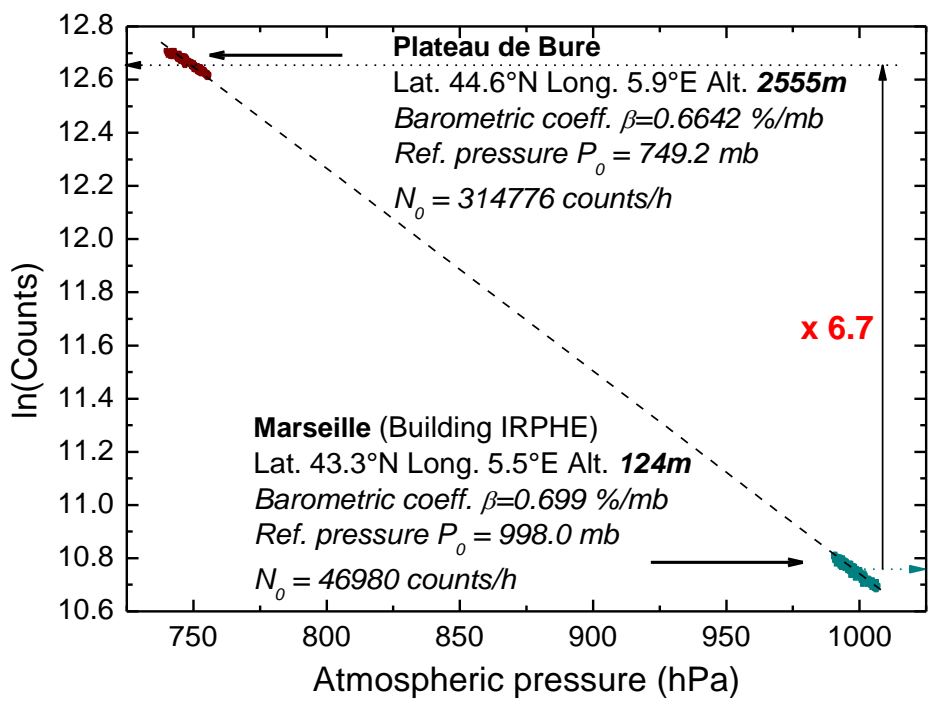

**Figure 4**

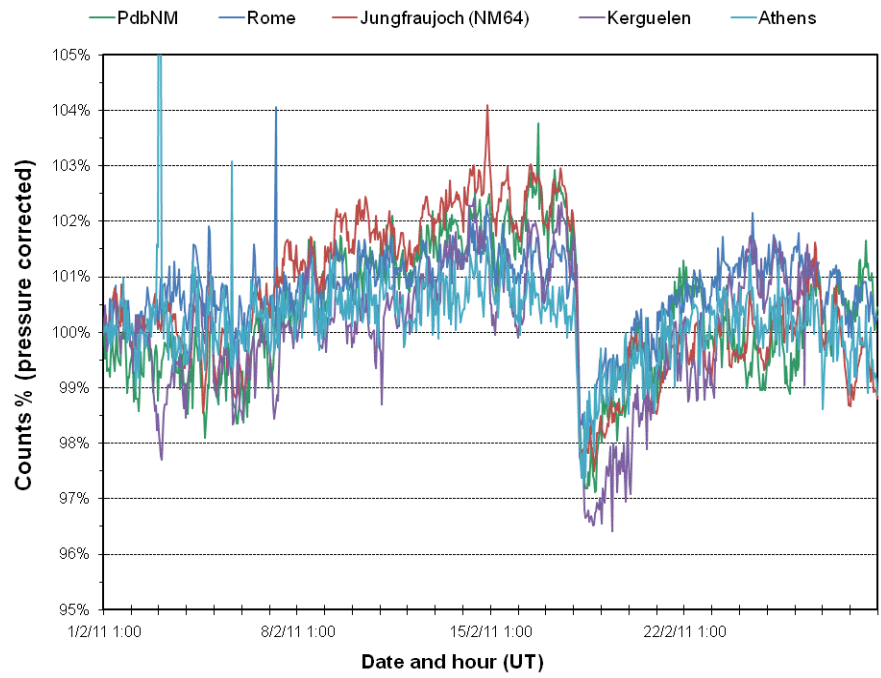

Figure 5

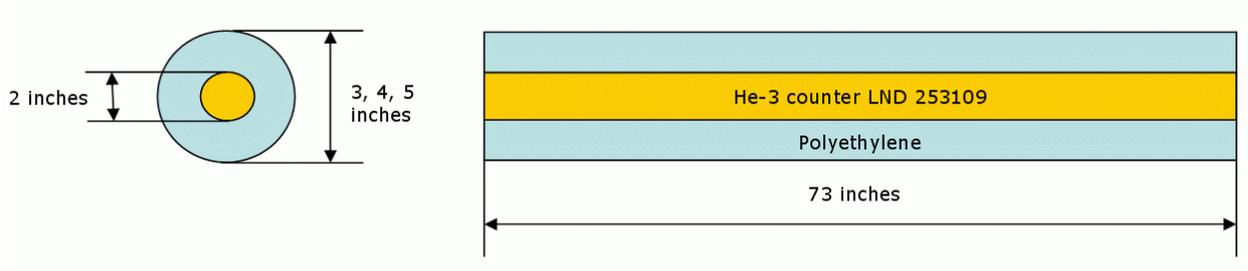

**Figure 6**

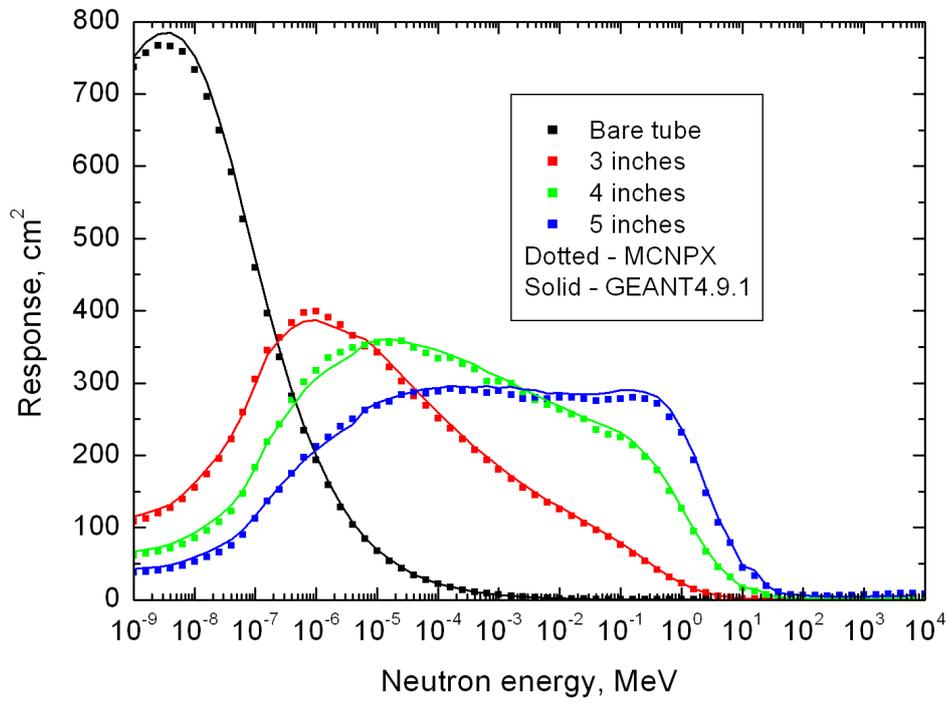

**Figure 7**

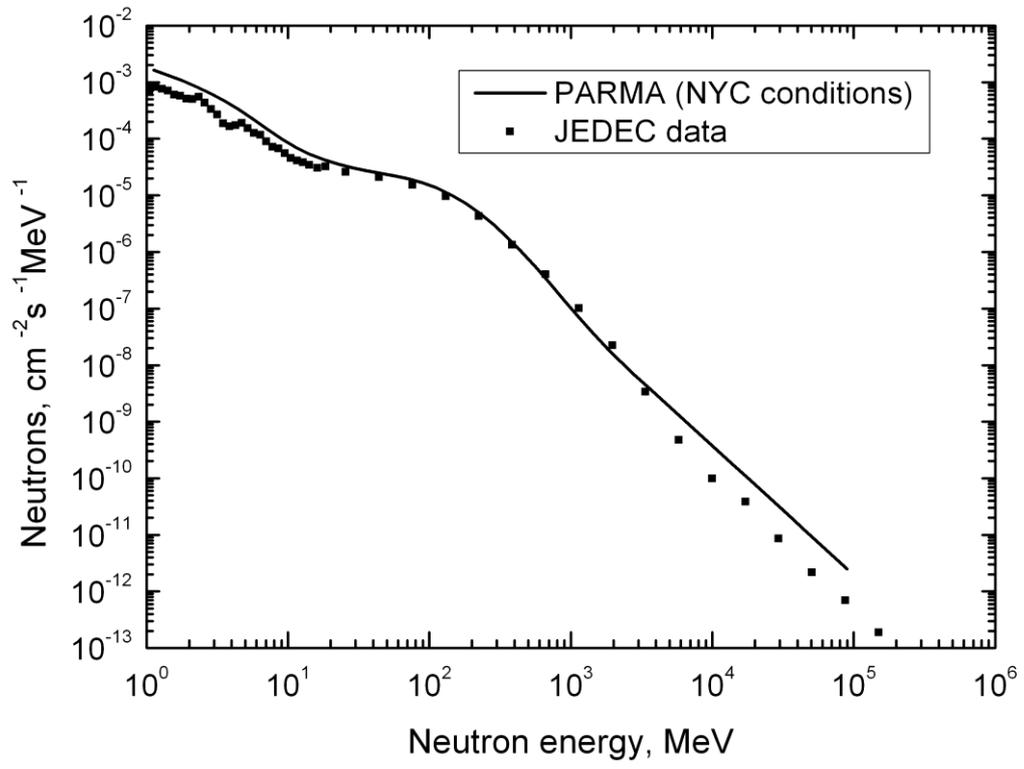

**Figure 8**

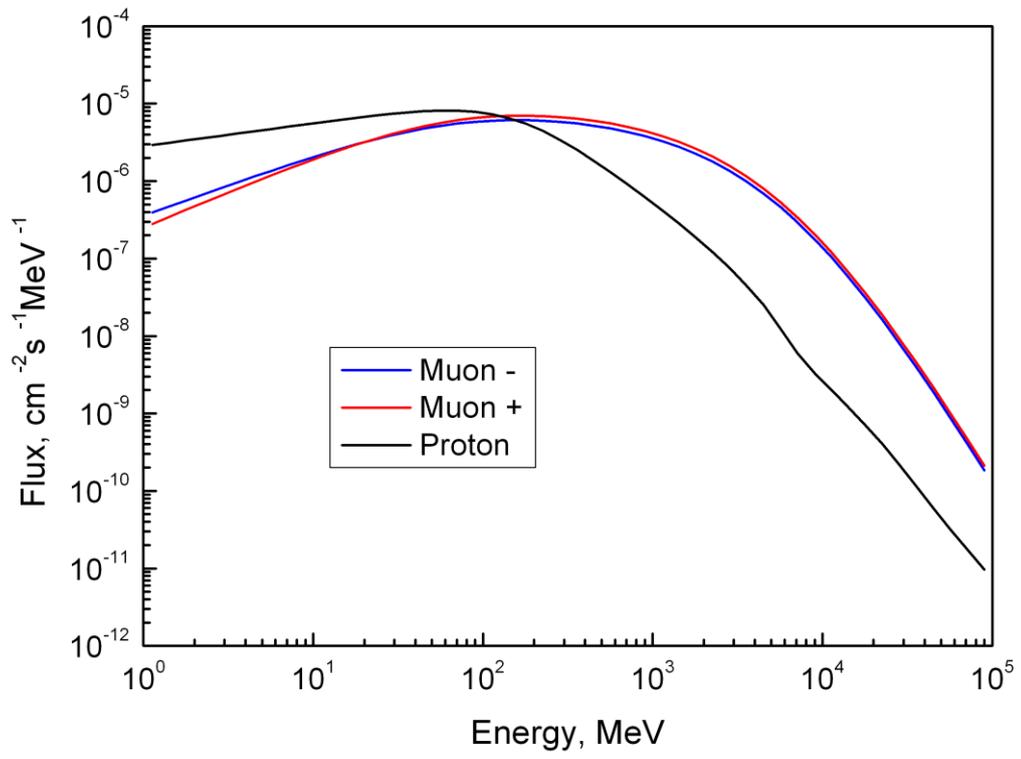

**Figure 9**

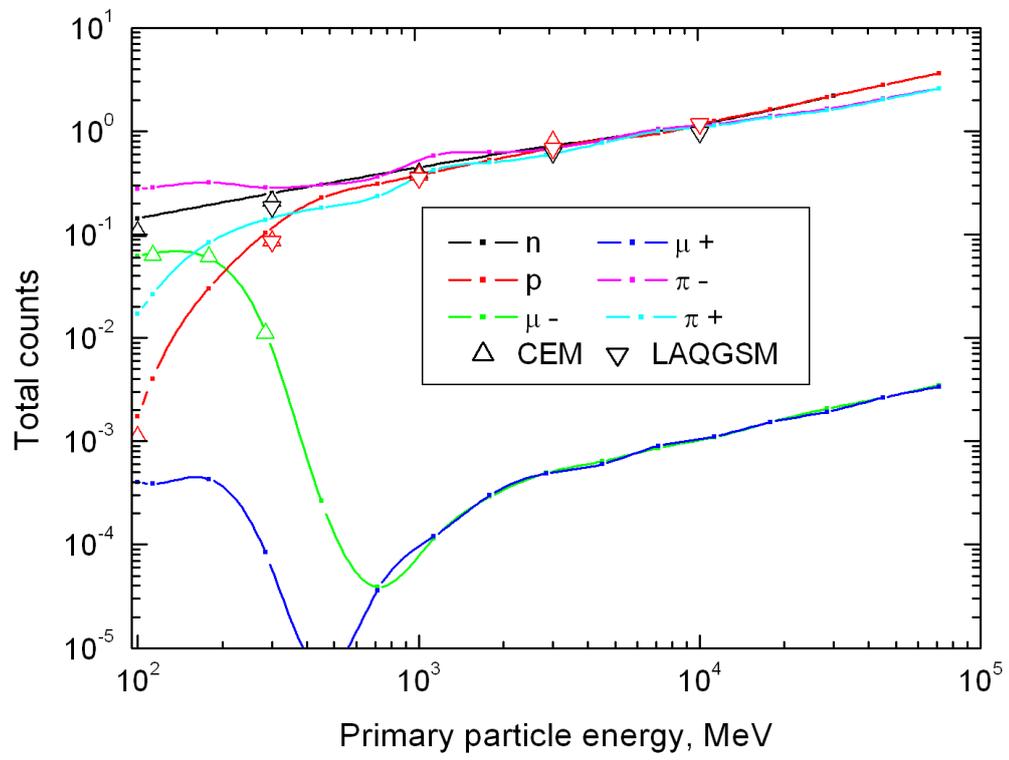

**Figure 10**

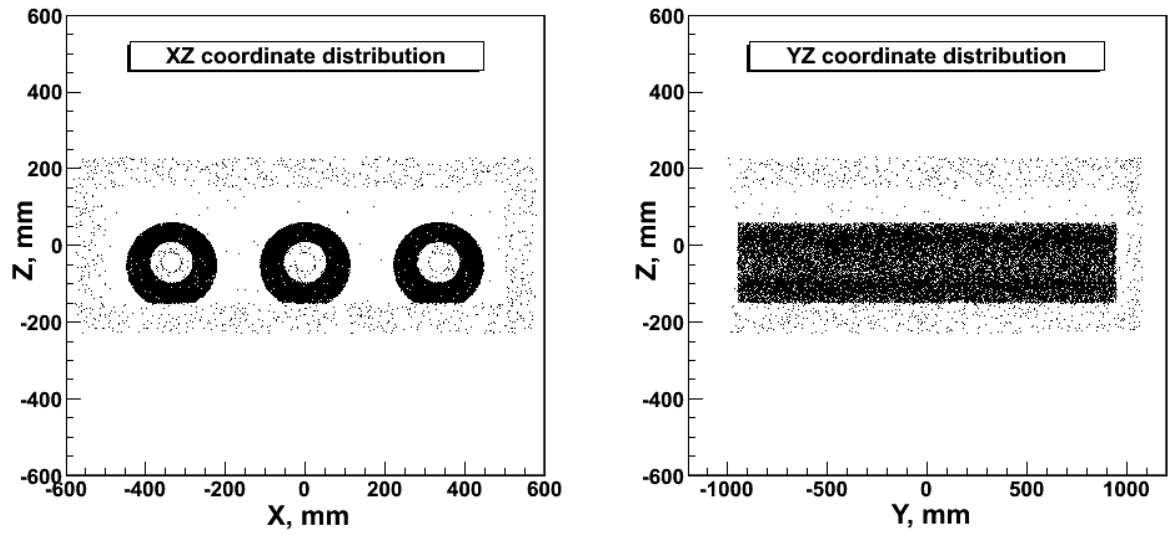

**Figure 11**

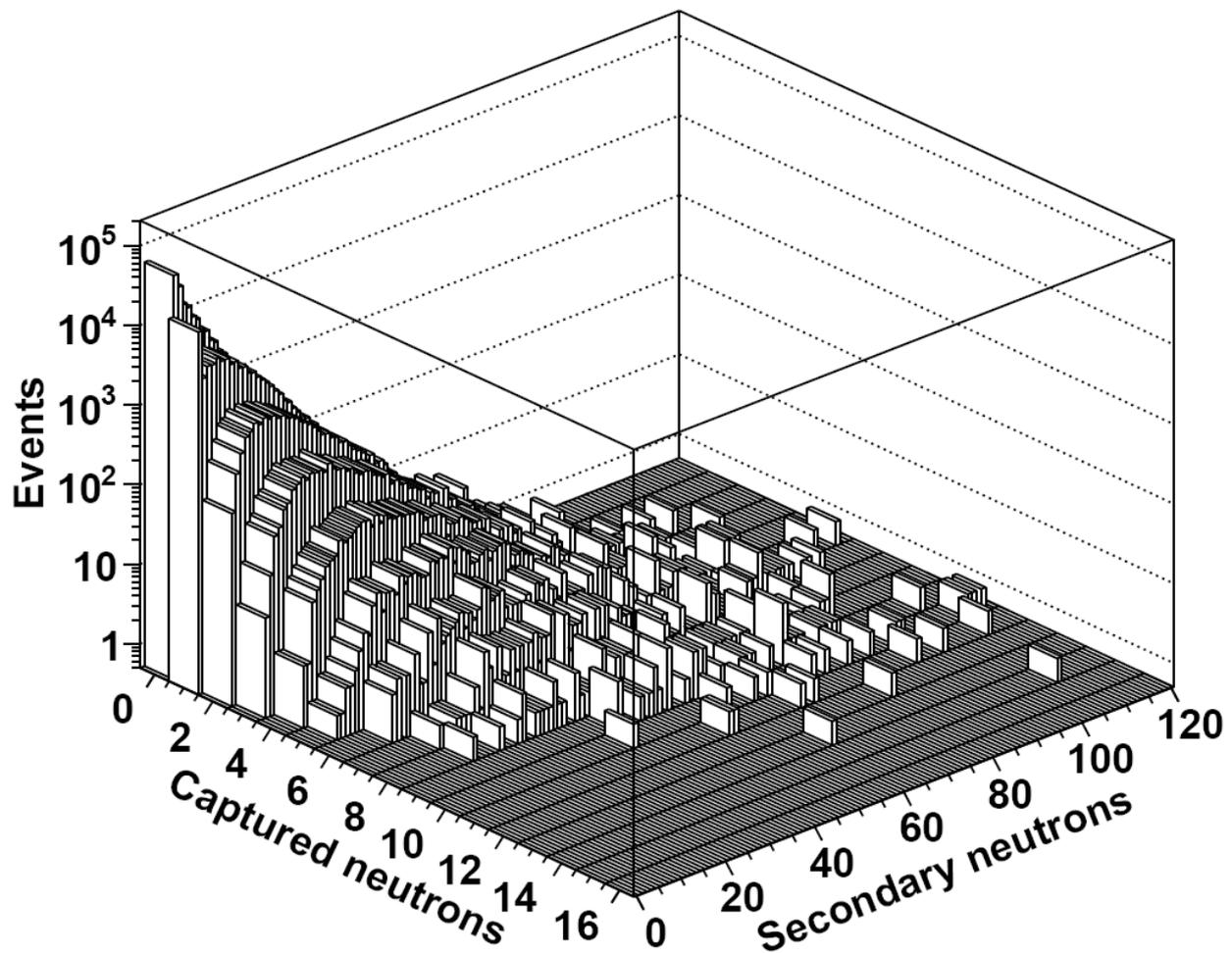

**Figure 12**

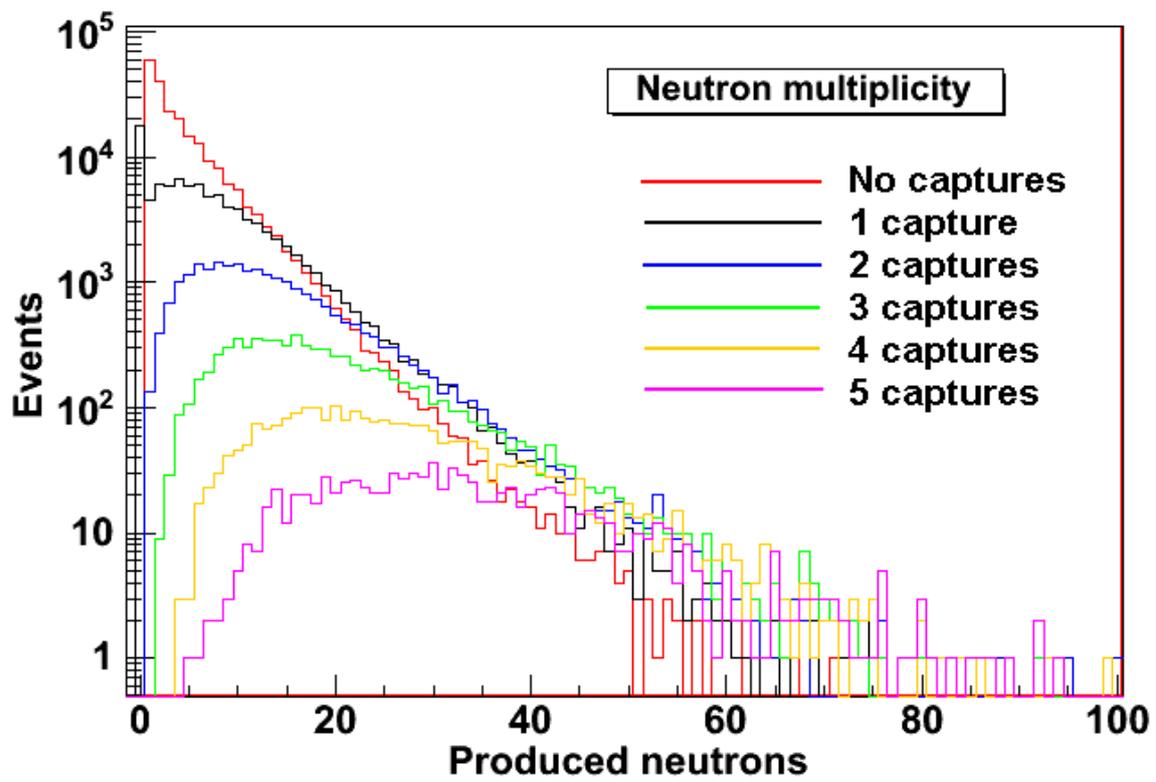

**Figure 13**

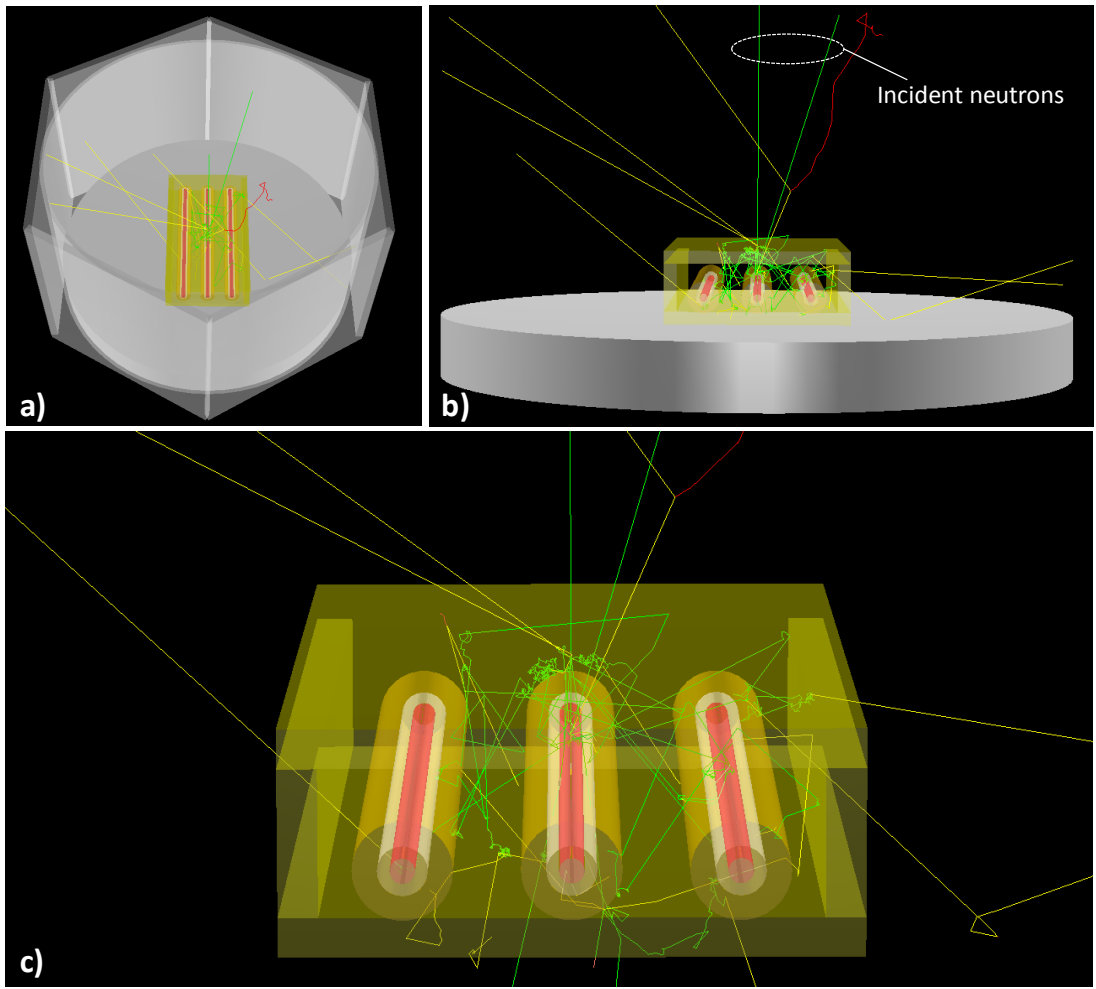

**Figure 14**

| **Neutron process** | **Energy** | **GEANT4 model** | **Dataset** |
|---|---|---|---|
| Elastic | < 4 eV | G4NeutronHPThermalScattering | G4N…HPThermalScatteringData |
|  | < 20 MeV | G4NeutronHPElastic | G4NeutronHPElasticData |
|  | > 20 MeV | G4LElastic | - |
| Inelastic | < 20 MeV | G4NeutronHPInelastic | G4NeutronHPInelasticData |
|  | [20 MeV, 10 GeV] | G4BinaryCascade | - |
|  | [10 GeV, 25 GeV] | G4LENeutronInelastic | - |
|  | [12 GeV, 100 TeV] | QGSP | - |
| Fission | < 20 MeV | G4NeutronHPFission | G4NeutronHPFissionData |
|  | > 20 MeV | G4LFission | - |
| Caption | < 20 MeV | G4NeutronHPCapture | G4NeutronHPCaptureData |
|  | > 20 MeV | G4LCapture | - |

**Table 1**

| $E_{prim}$, MeV | Detection efficiency | Total counts | $N_{sec} / N_{prim}$ |
|---|---|---|---|
| 100 | 0.112 | 0.143 | 1.786 |
| 300 | 0.157 | 0.251 | 3.589 |
| 1000 | 0.206 | 0.446 | 6.700 |
| 3000 | 0.236 | 0.721 | 10.950 |
| 10000 | 0.275 | 1.164 | 17.680 |

**Table 2**

|  | Contributions from the different multiplicities of neutron captures, % | | | | | | | | |
|---|---|---|---|---|---|---|---|---|---|
|  | 1 | 2 | 3 | 4 | 5 | 6 | 7 | 8 | 9 |
| Neutron | 51.1 | 26.0 | 12.5 | 5.5 | 2.6 | 1.2 | 0.6 | 0.3 | 0.1 |
| Proton | 24.2 | 23.7 | 17.2 | 11.5 | 7.3 | 5.0 | 3.2 | 2.1 | 1.4 |
| Muon─ | 88.2 | 10.1 | 1.1 | 0.2 | 0.2 | 0.1 | 0 | 0 | 0 |

**Table 3**

| $E_{prim}$, MeV | Protons | | | Negative muons | | |
|---|---|---|---|---|---|---|
| | Detection efficiency | Total counts | $N_{sec}/N_{prim}$ | Detection efficiency | Total counts | $N_{sec}/N_{prim}$ |
| 113 | 0.0038 | 0.0040 | 0.093 | 0.0614 | 0.0657 | 0.739 |
| 179 | 0.0248 | 0.0301 | 0.549 | 0.0556 | 0.0594 | 0.661 |
| 284 | 0.0743 | 0.1031 | 1.642 | 0.0108 | 0.0115 | 0.153 |
| 450 | 0.1424 | 0.2252 | 3.447 | 0.00024 | 0.00026 | 0.00285 |
| 713 | 0.1714 | 0.3088 | 4.829 | $3.9 \cdot 10^{-5}$ | $3.9 \cdot 10^{-5}$ | 0.00068 |
| 1130 | 0.1941 | 0.4028 | 6.338 | 0.00010 | 0.00011 | 0.00194 |
| … | … | … | … | … | … | … |
| 45000 | 0.3005 | 2.787 | 42.18 | 0.0019 | 0.0026 | 0.0374 |
| 71300 | 0.3075 | 3.633 | 54.95 | 0.0025 | 0.0035 | 0.0474 |

**Table 4**

| Location / Particles | Marseille (sea-level) | ASTEP (elev. 2555m) |
|---|---|---|
| Neutrons | 83.2 ÷ 83.4 % | 83.4 ÷ 86.6 % |
| Protons | 6.2 ÷ 8.6 % | 9.0 ÷ 13.2 % |
| Muons - | 6.0 ÷ 7.9 % | 2.3 ÷ 3.0 % |
| Gamma | 2.0 ÷ 2.7 % | 1.1 ÷ 1.4 % |

**Table 5**